\documentclass[aps,pre,twocolumn,superscriptaddress]{revtex4-1}
\usepackage{graphicx}
\usepackage{epstopdf}
\usepackage{bm}
\usepackage{amsmath}
\usepackage{verbatim}
\usepackage{amssymb}
\usepackage{listings}
\usepackage{color}
\definecolor{darkblue}{rgb}{0,0,0.6}
\usepackage{here}
\usepackage[colorlinks,linkcolor=darkblue,citecolor=darkblue,urlcolor=darkblue]{hyperref}
\renewcommand\vec[1]{\boldsymbol{#1}}

\newcommand\Fig[1]{Fig.~\ref{#1}}
\newcommand\Sec[1]{Sec.~\ref{#1}}

\renewcommand\vec[1]{\boldsymbol{#1}}
\newcommand\eq[1]{Eq.~(\ref{#1})}

\newcommand\bigang[1]{\left\langle #1 \right\rangle}
\newcommand\bigpar[1]{\left( #1 \right)}

\begin{document}

\title{Relaxation dynamics of non-Brownian spheres below jamming}

\author{Yoshihiko Nishikawa}


\affiliation{Laboratoire Charles Coulomb (L2C), Universit\'e de Montpellier, CNRS, 34095 Montpellier, France}
\author{Atsushi Ikeda}

\affiliation{Graduate School of Arts and Sciences, University of Tokyo, Komaba, Tokyo 153-8902, Japan}

\author{Ludovic Berthier}

\affiliation{Laboratoire Charles Coulomb (L2C), Universit\'e de Montpellier, CNRS, 34095 Montpellier, France}

\affiliation{Department of Chemistry, University of Cambridge, Lensfield Road, Cambridge CB2 1EW, United Kingdom}

\date{\today}

\begin{abstract}
We numerically study the relaxation dynamics and associated criticality of non-Brownian frictionless soft spheres below jamming in spatial dimensions $d=2$, $3$, $4$, and $8$, and in the mean-field Mari-Kurchan model. We discover non-trivial finite-size and volume fraction dependences of the relaxation time associated to the relaxation of unjammed packings. In particular, the relaxation time is shown to diverge logarithmically with system size at any density below jamming, and no critical exponent can characterise its behaviour approaching jamming. In mean-field, the relaxation time is instead well-defined: it diverges at jamming with a critical exponent that we determine numerically and differs from an earlier mean-field prediction. We rationalise the finite $d$ logarithmic divergence using an extreme-value statistics argument in which the relaxation time is dominated by the most connected region of the system. The same argument shows that the earlier proposition that relaxation dynamics and shear viscosity are directly related breaks down in large systems. The shear viscosity of non-Brownian packings is well-defined in all $d$ in the thermodynamic limit, but large finite-size effects plague its measurement close to jamming. 
\end{abstract}

\maketitle

\section{Introduction}
\label{sec:introduction}
The jamming transition is an athermal critical phase transition between a fluid at low density and a disordered solid at large density~\cite{Liu1998,OHern2002,OHern2003}. Exactly at jamming, particles form a rigid network and the number of contacts $z$ between particles is equal to $z_\mathrm c=2d$, where $d$ is the spatial dimension~\cite{OHern2003,Goodrich2012}. The contact forces and distances between particles follow power-law distributions with non-trivial exponents~\cite{Wyart2012,Charbonneau2012,Lerner2013,Charbonneau2015}. Above jamming, the excess contact number $z-2d$, mechanical modulus, 
and vibrational properties also exhibit non-trivial critical behavior~\cite{OHern2003,Wyart2005,Goodrich2012,Charbonneau2015}. The measured critical exponents agree with predictions from mean-field 
theories~\cite{Wyart2005,Charbonneau2014b,DeGiuli2014} down to $d=2$, suggested to be the upper critical dimension for jamming. 

By contrast to these well-understood results, the criticality of systems approaching jamming from the unjammed phase is still under intense scrutiny~\cite{Heussinger2009,Heussinger2010,Andreotti2012,Ikeda2013,Ikeda2013b,Vagberg2014,Trulsson2015,Atkinson2016,Hexner2018,Hexner2019,Arceri2020}. When jamming is approached from below, the shear viscosity exhibits a critical divergence~\cite{Olsson2007,mewis2012colloidal,Forterre2008,Boyer2011,Lerner2012,Olsson2011,Andreotti2012,Kawasaki2015,Olsson2020}. Similarly, in both sheared and isotropic systems, the relaxation dynamics slows down approaching jamming~\cite{Durian1995,Hatano2009,Olsson2015,Olsson2019,Ikeda2020,Saitoh2020,ikeda2020note}. In addition, both quantities have been linked to the frequency $\omega_{\min}$ of the softest vibrational mode characterising the structure of unjammed packings~\cite{Lerner2012}. Recent numerical simulations and theoretical arguments suggested that the relaxation time $\tau$, the viscosity $\eta$, and the inverse squared frequency $1/\omega_{\min}^2$, all follow the same critical behavior
\begin{equation}
\label{eq:critical_div}
\eta \sim \tau \sim 1/\omega_{\min}^2 \sim (\Delta z)^{-\beta},
\end{equation}
with $\Delta z \equiv 2d - z > 0$ now defining the shortage of contacts from the isostatic limit. 

If correct, the physical content of Eq.~(\ref{eq:critical_div}) is remarkable since it connects an important physical quantity, the shear viscosity of non-Brownian suspensions, to a diverging relaxation time, thus connecting the viscosity divergence to some form of dynamic slowing down. Further, the connection with the slowest mode of the vibrational density of states would then finally relate two macroscopic quantities (the relaxation time and the viscosity) to the detailed microstructure of unjammed packings, whose geometry controls the vibrational density of states~\cite{Lerner2012,Ikeda2020}. Physically, Eq.~(\ref{eq:critical_div}) follows from the numerical observation that the field of particle displacements under shear or during relaxational dynamics both have a very strong overlap with the spatial structure of the slowest vibrational mode~\cite{Lerner2012,Ikeda2020}.     

Despite these recent developments, both the validity and physical content of  
Eq.~(\ref{eq:critical_div}) remain open issues. There is no consensus yet on the value of the exponent $\beta$ and its physical origin. Theoretical arguments developed in Refs.~\cite{Lerner2012,Lerner2012a,DeGiuli2015,ikeda2020note} result in a quantitative prediction, $\beta \simeq 3.41$, which is directly linked to the values of known critical exponents for the jamming transition. This suggests that $\beta$ should also be independent of the spatial dimension for $d \geq 2$. Recently, a mean-field study analysed the exponent $\beta$ using the perceptron model of jamming, with a different prediction, $\beta \simeq 2.55$~\cite{Hwang2020}, again expected to hold in any $d \geq 2$. However, several simulations showed that the exponent $\beta$ depends on $d$ in a surprising manner. The relaxation time in $d=2$ and $d=3$ models has $\beta \simeq 2.7$ and $\beta \simeq 3.3$, respectively. These values were measured from the relaxation dynamics starting from steady shear trajectories~\cite{Olsson2019}. The value $\beta \simeq 3.2$ is reported in both sheared and isotropic cases for $d=3$~\cite{Ikeda2020}, from relaxation dynamics as well. More diverse values $\beta \in [2.0 -2.8]$ are obtained by direct measurements of the shear viscosity, which is notoriously difficult to determine~\cite{Andreotti2012,Kawasaki2015,Olsson2007,Olsson2020,Nordstrom2010}. Therefore, the determination of a precise value, the dimensional dependence, the mean-field theoretical value of the dynamical exponent $\beta$, and its dependence on the chosen observable for its measurement, are all unresolved issues. 

In this paper, we numerically study the dynamics of unjammed packings with three main goals. (i) We analyse the relaxation dynamics over a broad range of spatial dimensions, $d=2$, 3, 4 and 8 to assess its dimensionality dependence. (ii) We analyse the mean-field Mari-Kurchan (MK) model in order to resolve the conflict between current simulations and the mean-field prediction from the perceptron model, and also in order to complete the understanding of the dimensional dependence of the relaxation dynamics by including the mean-field limit. (iii) We perform a careful analysis of finite-size effects for the relaxation dynamics and the shear viscosity to directly assess the validity of Eq.~(\ref{eq:critical_div}) over a broad range of conditions.   

We find that the relaxation time $\tau$ seems to behave differently in $d=2$ ($\beta \simeq 2.8$) and in $d=3$, 4 and 8 ($\beta \simeq 3.3$), and find that $\beta \simeq 3.3$ also holds in the mean-field MK model, which would indicate that only $d=2$ is distinct from all $d \geq 3$ up to the mean-field $d \to \infty$ limit. However, we also find that strong finite-size effects affect the relaxation dynamics, and offer numerical and theoretical evidences that in the thermodynamic limit $N \to \infty$, the relaxation time diverges logarithmically, $\tau(N) \sim \log N$. This finite-size effect is at play in finite dimensional models $d < \infty$, but not in the mean-field MK model. 
We conclude that $\beta$ cannot be obtained from the relaxation time $\tau$ in finite dimensions, and that $\tau \sim (\Delta z)^{-\beta}$ in Eq.~(\ref{eq:critical_div}) is in fact incorrect. We finally demonstrate that the shear viscosity $\eta$ is finite and decoupled from $\tau$ as $N \to \infty$, meaning that $\eta \sim \tau$ in Eq.~(\ref{eq:critical_div}) is also incorrect, but that its determination is affected by large finite-size effects. We identify the length scale that controls this finite-size effect and find that it diverges so rapidly near jamming that viscosity measurements free from finite-size effects are extremely difficult. Our results establish that $\beta \simeq 3.3$ in the mean-field limit on the one hand, and identify a serious difficulty in the determination of $\beta$ in finite dimensions on the other hand. For the moment, we can neither rule out nor confirm the possibility that the mean-field value $\beta \simeq 3.3$ holds in any dimension $d \geq 2$ for the shear viscosity. 
 
The paper is organized as follows. In \Sec{sec:model_method}, 
we introduce our models of non-Brownian frictionless particles in finite dimensions and the Mari-Kurchan model, as well as important physical quantities. We present results for the relaxation time $\tau$ for all models in \Sec{sec:dyn_exp}. We discuss finite-size effects and their physical origin in \Sec{sec:log_growth}. In \Sec{sec:shear}, we present results for the shear viscosity, implying the breakdown of Eq.~(\ref{eq:critical_div}). In \Sec{sec:summary}, we discuss our results. 

\section{Models and methods}

\label{sec:model_method}

We study harmonic spheres defined by the following pair interactions
\begin{equation}
E = \frac\epsilon2 \sum_{i < j} \bigpar{1 - \frac{r_{ij}}{\sigma_{ij}}}^2
\Theta\bigpar{1 - \frac{r_{ij}}{\sigma_{ij}}},
\label{eq:energy}
\end{equation}
where $r_{ij} = \left|\vec r_i - \vec r_j - \vec A_{ij}\right|$ with $\vec r_i$ the 
position of particle $i$, $\sigma_{ij} = \sigma_i + \sigma_j$ with $\sigma_i$ 
the radius of particle $i$, and $\Theta\bigpar{x}$ is the Heaviside step function. We choose $\bigpar{\vec A_{ij}} = \mathrm{unif}(0, L)$ 
with $L$ the linear length of the system to analyse the mean-field Mari-Kurchan (MK) model~\cite{Kraichnan1962,Mari2009,Mari2011,Charbonneau2014}, while $\vec A_{ij} = \vec 0$ for finite dimensional models in $d$ dimensions. The physical idea behind the long-range random shifts in the MK model is that particles live in a finite-dimensional space, but interact with each other as if they were on a random graph with only a very small number of local loops and thus a tree-like structure, similar to the Bethe lattice~\cite{Kraichnan1962}. This geometry decreases the role of long-range, multi-body correlations and this makes the MK model behave as a particle model in the large dimensional limit with mean-field behaviour. The mean-field MK model may thus be closer to physical particle models than the perceptron analysed in Ref.~\cite{Hwang2020}. We study the MK model in three dimensions throughout this paper.

We use a $50:50$ binary mixture for the two dimensional case with $\sigma_\mathrm{large}/\sigma_\mathrm{small} = 1.4$, and monodisperse models for the other cases including the MK model. The boundary conditions are periodic in all directions for the isotropic case, and Lees-Edwards conditions~\cite{Lees1972} are used when simple shear is applied. We denote $N$ the number of particles, which we vary systematically in our simulations. The simulations are performed in a hypercubic box of linear size $L \sim N^{1/d}$. We denote $\phi$ the volume fraction of the system, which corresponds to the volume occupied by the particles divided by $V=L^d$.   

In the isotropic case, we study the relaxation dynamics of the model by studying overdamped Langevin dynamics at zero-temperature:
\begin{equation}
\label{eq:eom}
\zeta\frac{\partial \vec r_i}{\partial t} = - \frac{\partial E}{\partial \vec r_i},
\end{equation}
starting from a given completely random configuration in the high-temperature limit at a fixed volume fraction, where $\zeta$ is the damping coefficient (This could be realised experimentally by imposing a very large shear rate). In this model, the energy dissipation occurs between particles and the fixed background, rather than the local contacts between particles as in the model introduced by Durian \cite{Durian1995}. Whereas the local contacts of particles should become more relevant approaching jamming, we use the model for its simplicity and because the dynamical criticality remains the same in the two dissipation models \cite{Olsson2015}. The typical number of samples at each volume fraction studied in this paper is $10^3$. The time unit of the dynamics is $\tau_0 = \zeta \sigma_\mathrm{small}^2 / \epsilon$ for the binary mixture and $\tau_0 = \zeta \sigma^2 / \epsilon$ for the monodisperse model. We express timescales in units of $\tau_0$. Similarly, $\sigma_\mathrm{small}$ or $\sigma$ are the unit length. The equation of motions are solved by the simple Euler method. During the overdamped dynamics, the energy asymptotically decays exponentially, $E \sim \exp(-t / \tau)$. We thus define 
\begin{equation}
\label{eq:relax}
\tau = -\bigpar{\frac{d(\log E)}{dt}}^{-1}
\end{equation} 
as the relaxation time when the energy density reaches the value $E / N = 10^{-18}$. Before entering the asymptotic exponential regime, the local relaxation time \eq{eq:relax} measured for each realisation typically grows with time before converging to its final asymptotic value. The threshold $E / N = 10^{-18}$ is small enough that the local relaxation time converges for all the system sizes studied in this paper. Since the relaxation time defined in this manner is computed in the final stages of the energy relaxation towards an unjammed configuration, it corresponds to the slowest timescale characterising the relaxation process.
We stress that the relaxation time $\tau$ is different from a relaxation time measured in an equilibrium context, which usually quantifies the time decay of a microscopic correlation function. Here, $\tau$ describes the timescale needed to reach the final unjammed state starting from a given initial condition with finite overlaps. 

Since we stop the overdamped dynamics when the energy density $E / N = 10^{-18} > 0$, many particles in the final configurations have finite (but very small) overlaps with their neighbors and are thus subject to finite forces (which are vanishing exponentially with time and are zero at $t=\infty$). To characterize each final configuration, we count the contact number $z$ between particles. 
Before computing the contact number, we iteratively remove rattler particles that have less than $d$ contacts until all remaining particles have more than $d$ contacts~\cite{Lerner2012,Olsson2015}. The volume fraction after removing rattlers becomes slightly smaller.
We also compute the spatial correlation function of the forces in the final configurations
\begin{equation}
C_\mathrm{force}(r) = \frac{\bigang{\frac1N\sum_{i<j} \vec f_i \cdot \vec f_j \delta\bigpar{r - r_{ij}}}}
{\bigang{\frac1N\sum_{i} \vec f_i \cdot \vec f_i}}.
\label{eq:corr_func}
\end{equation}
Since the total momentum is zero in the overdamped dynamics \eq{eq:energy}, the correlation function $C_\mathrm{force}(r)$ has an anticorrelation, meaning that $C_\mathrm{force}(r)$ becomes negative at some $r > 0$. We define the typical correlation length $\xi_\mathrm{force}$ for force-force correlations as the distance where $C_\mathrm{force}(\xi_\mathrm{force}) = 0$ for the first time. During the relaxation, the correlation length grows and converges to the final asymptotic value with the same time scale as the local relaxation time \eq{eq:relax} converges.

To visualize real-space structures of physical quantities, we define the coarse-grained field $\hat O({\vec r})$ of an observable $O_i$ as follows: 
\begin{equation}
\hat O({\vec r}) = \frac{\sum_i W({\vec r} - \vec r_i) O_i}{\sum_i W({\vec r} - \vec r_i)},
\label{eq:cgfield}
\end{equation}
where we introduce a Gaussian window function of typical width $\sigma_{\rm filter}$
\begin{equation}
W(\vec r) = \left\{\begin{array}{ll}
e^{-\vec r^2 / (2 \sigma_\mathrm{filter}^2)}
- e^{-r_\mathrm{c}^2 / (2 \sigma_\mathrm{filter}^2)} &, \, |\vec r| < r_\mathrm{c},\\
0 &, \, |\vec r| > r_\mathrm{c},
\end{array}\right.
\end{equation}
with $\vec r_i$ the position of particle $i$.
Here, for example, $O_i = |\vec f_i|$ for the force, $O_i = z_i$ for the contact number,
where $\vec f_i$ and $z_i$ are the force acting on particle $i$ and the contact number of particle $i$, respectively. For illustrative purposes, we choose $r_\mathrm{c} = 30$ and $\sigma_\mathrm{filter} = 4$. The values of $\sigma_\mathrm{filter}$ and $r_\mathrm{c}$ do not change the visualization qualitatively unless they are small enough compared to the box size.

To analyse shear rheology, we study both the steady state properties and the relaxation dynamics of the binary mixture model in two dimensions.
To achieve the steady sheared state, we solve the equations of motion
\begin{equation}
\zeta \left(\frac{\partial \vec r_i}{\partial t} - \dot \gamma y_i \vec e_x\right) 
= - \frac{\partial E}{\partial \vec r_i} 
\end{equation}
with Lees-Edwards boundary conditions~\cite{Lees1972} in $y$ direction, where $\dot \gamma$ is the shear rate, $y_i$ is the $y$-coordinate of particle $i$, $\zeta$ is the damping coefficient, and $\vec e_x$ represents the unit vector parallel to $x$ direction. The time unit of the dynamics is again $\tau_0 = \zeta \sigma_\mathrm{small}^2 / \epsilon$. The shear viscosity $\eta$ and its pressure analog $\eta_p$ are defined as the shear 
stress $\sigma$ and the pressure $p$ divided by the shear rate $\dot \gamma$, respectively.
The shear stress $\sigma$ and the pressure $p$ are computed using the following formulae:
\begin{align}
&\sigma = \frac1N \sum_{i < j} \frac{x_{ij} y_{ij}}{r_{ij}} \frac{\partial E}{\partial r_{ij}},\\
&p = \frac1N \sum_{i < j} \frac{y^2_{ij}}{r_{ij}} \frac{\partial E}{\partial r_{ij}},
\end{align}
where $r_{ij} = |\vec r_i - \vec r_j|$, $x_{ij} = x_i - x_j$, and $y_{ij} = y_i - y_j$.
Starting from uniformly random configurations, we monitor the energy, the shear 
stress and the pressure to confirm the convergence to the steady state after the shear strain $\gamma = \dot \gamma t$ reaches $10$. In order to obtain the shear viscosity $\eta$ and $\eta_p$ in the Newtonian regime, we set $\dot \gamma = 10^{-6}$ and $10^{-7}$ for 
$\phi = 0.8$ and $0.83$, respectively. With these choices, $\eta$ and $\eta_p$ are independent of $\dot \gamma $ when it is lowered, consistent with Ref.~\cite{Olsson2015}.
We also studied the overdamped dynamics starting from sheared configurations taken during the steady state. To this end, we suddenly stop the shear and solve the equation of motion \eq{eq:eom}. We measure the relaxation time \eq{eq:relax} of the overdamped dynamics when $E / N = 10^{-18}$, exactly as in the isotropic case.

\section{Relaxation time in various spatial dimensions}

\label{sec:dyn_exp}

\subsection{Dynamical exponent in the Mari-Kurchan model}

We study the relaxation dynamics of the MK model. We solve Eq.~(\ref{eq:eom}) starting from random initial configurations. The volume fractions we studied in this model range from $\phi=1.42$ to $\phi = 1.46$. In the large-size limit $N \to \infty$, the system at the jamming transition shows an algebraic energy decay $E \sim t^{-\alpha}$. An algebraic decay is still observed in finite systems at short time with $\alpha \lesssim 1$ when the packing fraction 
is close enough to jamming, but the decay crosses over to being exponential at large times. In this regime, we compute the relaxation time $\tau$ using \eq{eq:relax}. 

\begin{figure}
\includegraphics[width=\linewidth]{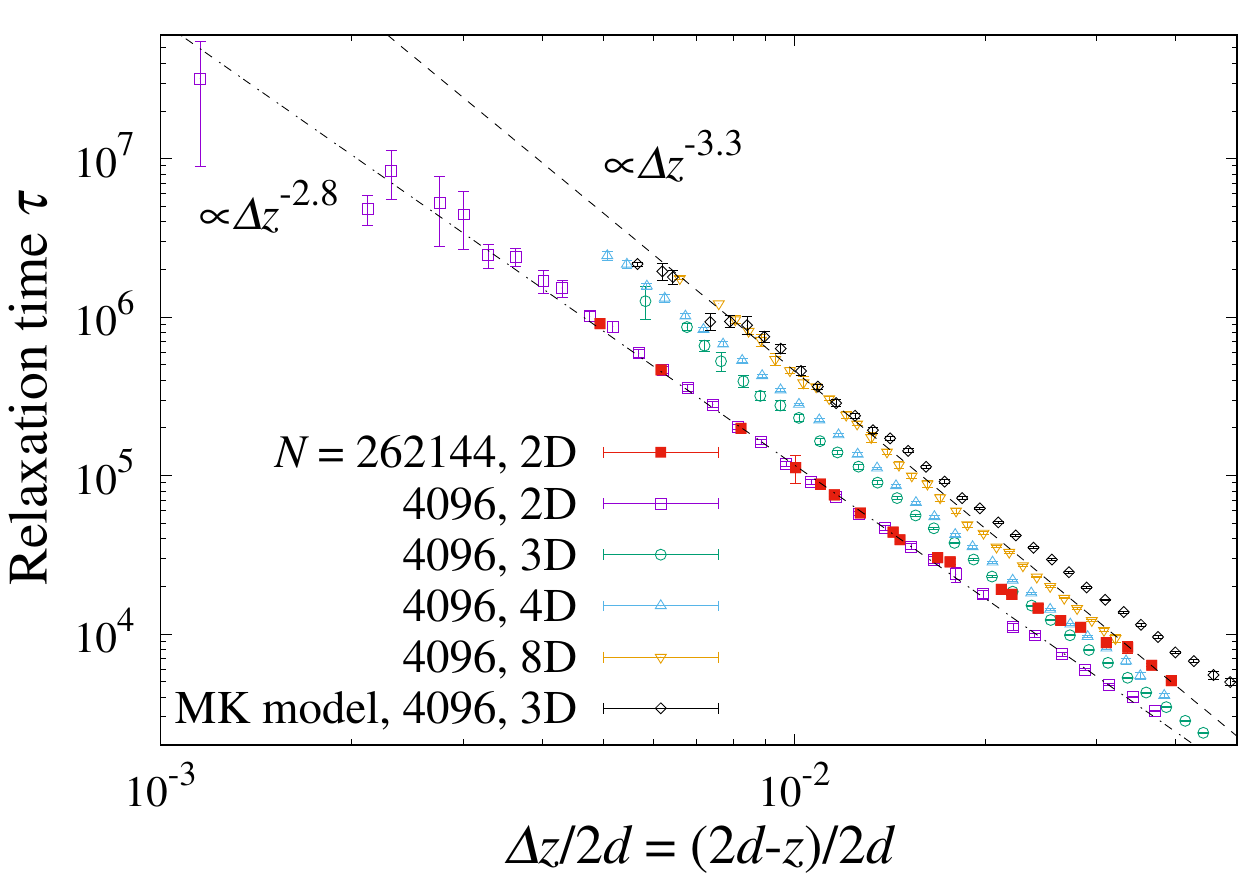}
\caption{Relaxation time $\tau$ as a function of $\Delta z = 2d - z$ in $d=2$, $3$, $4$, $8$ and in the mean-field MK model. For $d \geq 3$ and the MK model, 
$\beta \simeq 3.3$ at small $\Delta z$, whereas for $d=2$ $\beta \simeq 2.8$. Two system sizes are shown in $d=2$, suggesting that finite-size effects are large for large $\Delta z$.}
\label{fig:tau_vs_dz}
\end{figure}

In \Fig{fig:tau_vs_dz}, we plot $\tau$ against $\Delta z = 2d - z$, the shortage of the contact number in the final configuration. The contact number is calculated after removing 
the rattlers (see \Sec{sec:model_method}). Note that larger $\Delta z$ values correspond to lower packing fractions. The figure shows that $\tau$ cannot be fitted by a single power-law in the entire 
density region. If we focus on the lower density region $\Delta z / 2d \gtrsim 4 \times 10^{-2}$, 
the power-law fitting \eq{eq:critical_div} gives the exponent $\beta \simeq 2.6$. This is 
comparable with an estimated value for the relaxation time of the perceptron model 
$\beta = 2.55(15)$ \cite{Hwang2020}. However, the relaxation time at smaller $\Delta z$ 
grows more rapidly. It eventually crosses over to another power-law behavior with $\beta 
\simeq 3.3$ at $\Delta z / 2d \lesssim 10^{-2}$. Even if we evaluate the exponent for our 
data up to $\Delta z / 2d \simeq 3 \times 10^{-2}$ by fitting to \eq{eq:critical_div},
it yields $\beta \simeq 3.0$, significantly larger than the estimation in Ref.~\cite{Hwang2020}.
Therefore, we conclude that the dynamical exponent $\beta$ of the MK model is not equal to that of the perceptron model~\cite{Hwang2020}. Our best estimate is $\beta \simeq 3.3$, but larger systems and a more careful finite-size scaling analysis would perhaps modify this value by a small amount. Such an analysis would be useful to compare the mean-field value to the prediction $\beta \simeq 3.41$ obtained in Ref.~\cite{DeGiuli2015}. 

\subsection{Relaxation dynamics in finite dimensions}

In sufficiently high dimensions, we may expect the system to approach some mean-field behaviour. Since two distinct mean-field models (the perceptron and the MK models) exhibit different dynamical exponents, it is important to directly study the dimensional dependence of the behaviour of $\tau$. We study the relaxation dynamics of harmonic spheres from random configurations in $d=2$, $3$, $4$ and $8$. As for the MK model, we measure the relaxation time $\tau$ using \eq{eq:relax}, and plot $\tau$ against $\Delta z$ in \Fig{fig:tau_vs_dz}. The jamming densities are known to be $\phi_J \simeq 0.842$ for the two-dimensional model \cite{Vagberg2011}, and $\phi_J \simeq 0.65$, $0.46$, $0.078$ for the three-, four-, and eight-dimensional models \cite{Sartor2020}, respectively. We thus study the volume fractions in the range of $[0.8, 0.842]$, $[0.627, 0.634]$, $[0.445, 0.451]$, and $[0.074, 0.075]$ for two, three, four, and eight dimensional models for \Fig{fig:tau_vs_dz}. 

We find that the exponent $\beta \simeq 3.3$ works well in a wide range of dimensions, from $d=8$ down to $d = 3$, where the relaxation time seems to follow a power-law with the same exponent $\beta \simeq 3.3$ at small enough $\Delta z$. The results in $d=3$ are consistent with previous reports~\cite{Olsson2019,Ikeda2020}. We conclude that, over the range of system sizes used in Fig.~\ref{fig:tau_vs_dz}, harmonic spheres in $d \geq 3$ and the MK model follow a very similar behaviour as they approach jamming where $\tau \to \infty$ and $\Delta z \to 0$. This behaviour seems different from the perceptron model prediction. 
  
A hint that this may not be the final story is the behaviour revealed by the two-dimensional model, which does not share the same apparent value of the exponent $\beta$. The growth of the relaxation time is 
noticeably milder in two dimensions, and the fitting gives an exponent $\beta \simeq 2.8$. These results confirm the surprising dimensional dependence of the exponent reported recently~\cite{Olsson2019}. 

The dimensional dependence of the apparent exponent could suggest the existence of a characteristic dimension between $d=2$ and $d=3$ where the dynamic criticality changes qualitatively. This interpretation would also suggest that the critical dynamics is not fully determined by static exponents for jamming since the static exponents are independent of dimensionality down to $d=2$ ~\cite{OHern2003,Wyart2005,Goodrich2012,Charbonneau2015} as we mentioned in \Sec{sec:introduction}.

We will in fact suggest a very different interpretation, namely that the exponent $\beta$ characterising the growth of $\tau$ when jamming is approached is actually not well-defined in finite-dimensional models. As a result, the dimensionality dependence of its apparent value is immaterial, and the findings in Fig.~\ref{fig:tau_vs_dz} should not be taken as the sign of an emerging non-trivial upper critical dimension for the dynamics of non-Brownian suspensions.

We can see hints of this interpretation in Fig.~\ref{fig:tau_vs_dz} where the relaxation time data for the two-dimensional system with $N=4096$ and $N=262144$ seem to show a similar power-law behavior at small $\Delta z$, but appear to behave very differently away from the transition above $\Delta z \gtrsim 2 \times 10^{-2}$. This suggests that finite-size effects should be analysed more precisely, as we discuss in the following.

\section{Logarithmic divergence of the relaxation time}

\label{sec:log_growth}

Whereas the relaxation time $\tau$ in finite-dimensional models seems to follow a  power-law divergence towards $\Delta z \to 0$, it also seems to suffer from finite-size effects, which are seen already at modest volume fractions and relatively large $\Delta z$ away from jamming. 

\subsection{Logarithmic growth of the relaxation time}

\label{sec:lg_relax}

\begin{figure}
\includegraphics[width=8.5cm]{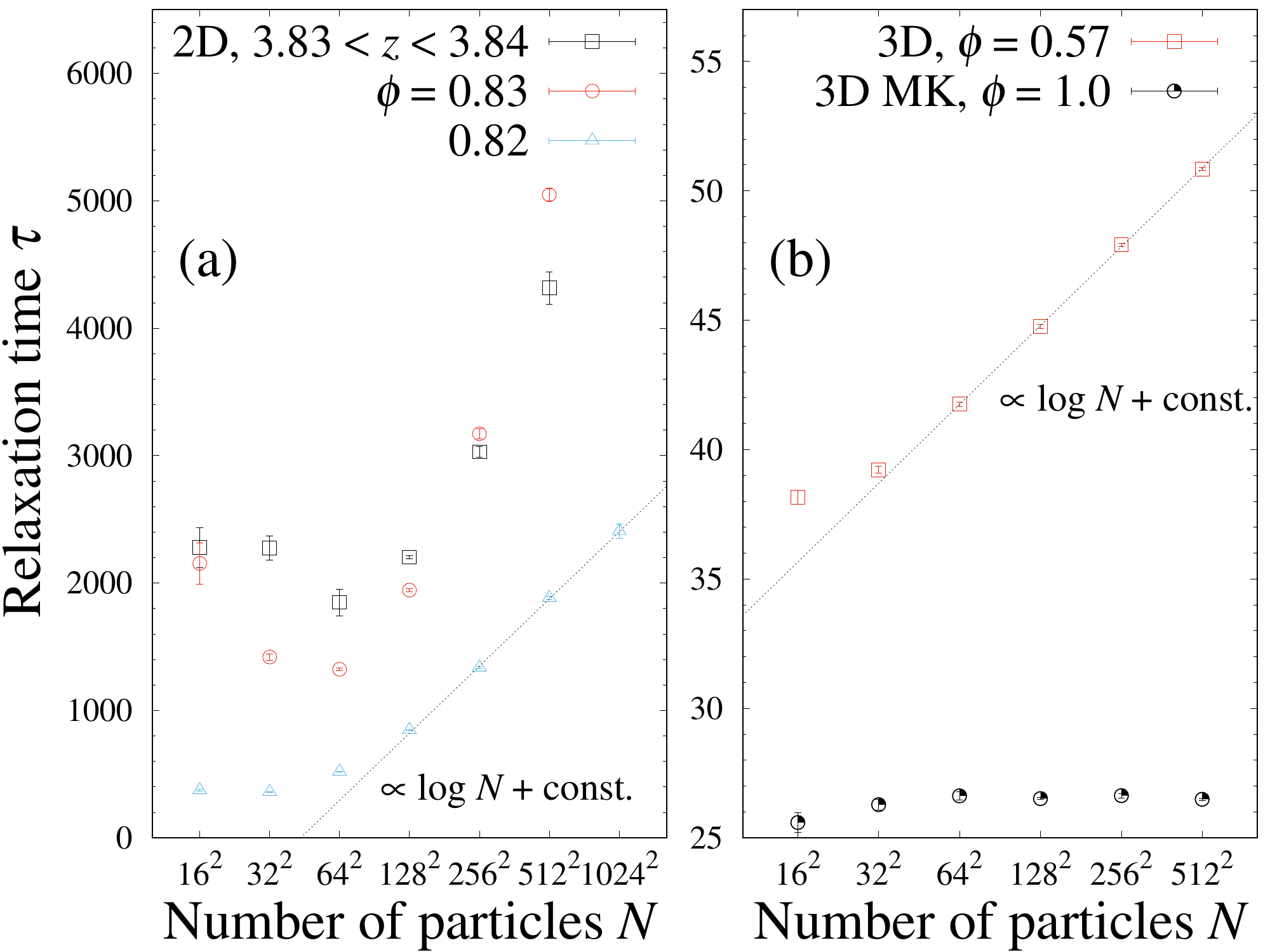}
\caption{(a) Relaxation time $\tau$ as a function of the particle number $N$ in $d=2$ averaged either at fixed $z$ or fixed volume fraction $\phi$. (b) Same plot in $d=3$ and for the Mari-Kurchan model at fixed $\phi$. A logarithmic growth $\tau \sim \log N$ in both $d=2$ and $d=3$, but is not observed in the MK model: $\tau(N\to\infty)$ is not defined in finite $d$.}
\label{fig:tau_vs_N}
\end{figure}

\Fig{fig:tau_vs_N}(a) shows the relaxation time $\tau$ as a function of the number of particles 
$N$ averaged over samples with contact number $z \in [3.83, 3.84]$, which corresponds to $\Delta z / 2d \in [4 \times 10^{-2}, 4.25 \times 10^{-2}]$, and with fixed volume fractions $\phi=0.82$ and $\phi= 0.83$ in two dimensions. 

In small systems, the contact number at fixed 
density has large sample-to-sample fluctuations. The relaxation time of small systems averaged at fixed density receives contributions from samples with a wide range of contact numbers, and this is expected to have larger finite-size effects compared to the case of an average at a fixed contact number. Indeed, the relaxation time averaged at fixed volume fraction $\phi = 0.83$ has a more pronounced dependence on $N$ below $N = 64^2$ than at fixed contact number.

At much larger $N$, on the other hand, the relaxation time $\tau$ grows similarly regardless of the averaging procedure. Crucially,  its asymptotic dependence is a logarithmic growth, $\tau(N) \sim \log N$. This unbounded logarithmic growth with the system size suggests that the relaxation time $\tau$ is not well-defined in the thermodynamic limit, as it diverges when $N \to \infty$ at a finite distance below the jamming transition. We find the slope becomes smaller as $\phi$ decreases and it would go to zero in the limit $\phi \to 0$, but the logarithmic growth should appear at any finite $\phi$ below jamming.

This finite-size effect is surprising since the system is not expected to have a diverging lengthscale in the entire unjammed phase, but to be critical at the jamming transition only. We may wonder whether two dimensions is a special dimension, as it is for instance for other types of equilibrium phase transitions~\cite{Mermin1966}. In \Fig{fig:tau_vs_N}(b), we show that the same logarithmic behaviour at large $N$ is in fact present for $d=3$ (here, at fixed volume fraction $\phi=0.57$). We expect the logarithmic behaviour to be a generic feature in any $d<\infty$ below jamming.

In the previous section, we discussed the value of an apparent critical exponent $\beta$ for the relaxation time. However, because the relaxation time is divergent in the entire unjammed phase, we have to carefully consider the physical meaning of these measurements. We emphasize that when the volume fraction is very close to the jamming transition, the logarithmic growth only appears when very large systems sizes are considered. For instance, \Fig{fig:tau_vs_dz} shows that the relaxation time at $N=262144$ and $N=4096$ nearly coincide for $\Delta z / (2d) \lesssim 2\times 10^{-2}$, where we determined an effective critical exponent. We observe that it becomes harder to detect the logarithmic growth closer to jamming, which presumably explains why this effect has not been detected before. This suggests the existence of a lengthscale $\xi(\phi)$ controlling the emergence of a logarithmic divergence, and $\xi(\phi)$ may be controlled by the distance to jamming, as we directly confirm below. 

Finally, \Fig{fig:tau_vs_N}(b) also shows the system size dependence of the relaxation time in the mean-field MK model. Clearly, the MK model is free from the logarithmic divergence at large $N$. Therefore, the measured dynamical exponent $\beta$ is well-defined even in the thermodynamic limit for the mean-field model. This suggests that the physical origin of the logarithmic growth is a finite dimensional effect affecting the definition and determination of $\beta$ when $d < \infty$. Numerical observations in dimensions $d=4$ and $8$ suggest that over a broad range of system sizes and volume fractions, these systems behave very similarly to the MK model, which may then serves as a useful guide for finite $d$ systems but only over a finite regime.   

\subsection{Relaxation with multiple islands}

\label{sec:island}

Because of the overdamped equation of motion, particle motion is 
directly controlled by the net forces acting on the particles. In particular, the displacement of each particle during the terminal relaxation should be proportional to the force acting on each particle in the final 
configuration~\cite{Ikeda2020}. This suggests that an analysis of the force field should shed light on the relaxation dynamics and provide insight on the physical origin of the logarithmic divergence of $\tau$. We focus on $d=2$, as larger system sizes can be simulated and visualisation is much easier. 

\begin{figure}
\includegraphics[width=8.5cm]{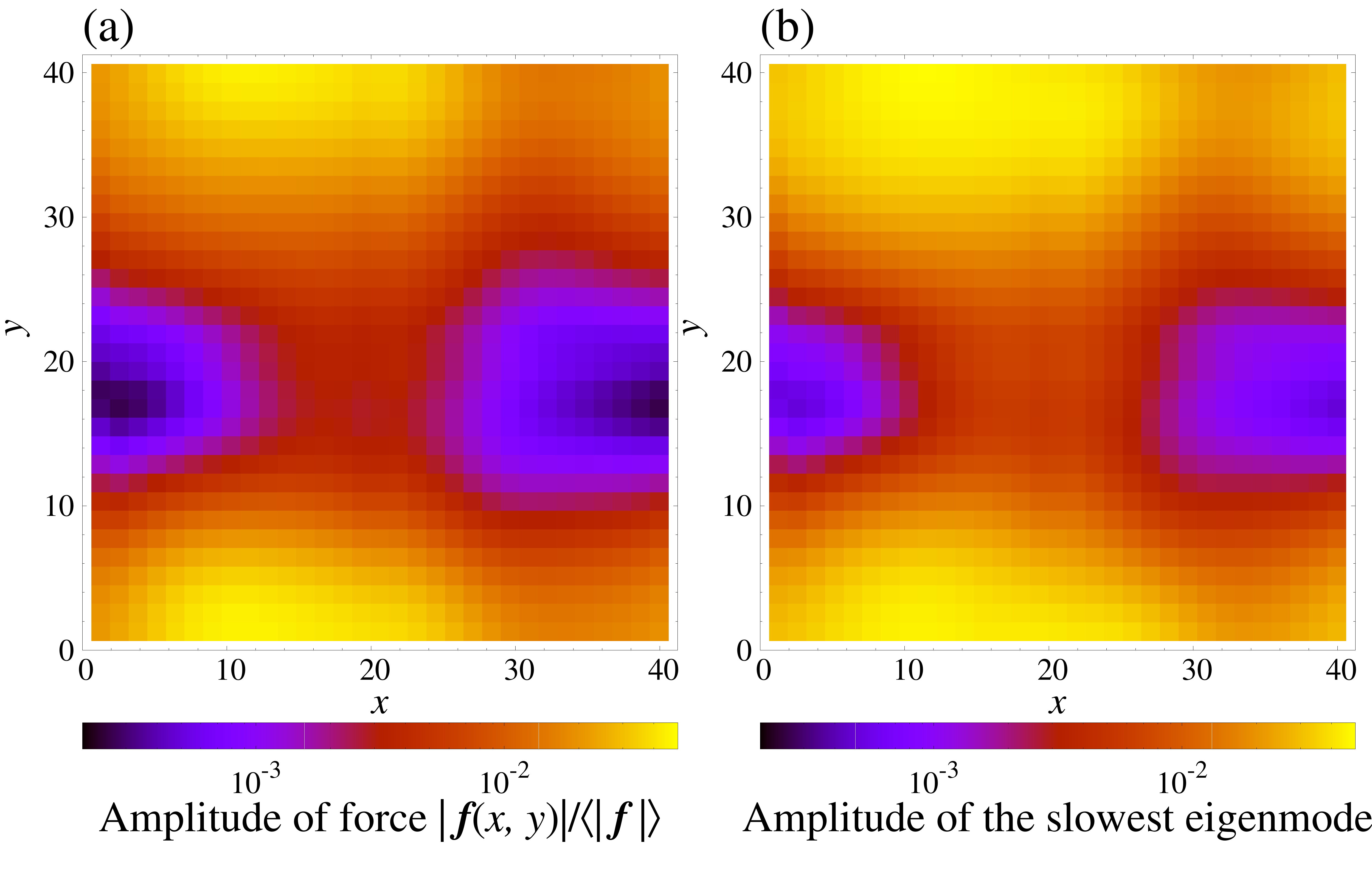}
\caption{Coarse-grained fields (see \eq{eq:cgfield}) for (a) the amplitude of the normalized force, and (b) the slowest eigenmode in a relaxed unjammed configuration at $\phi = 0.7$ with $N=1024$. Both fields are almost equivalent and seem correlated over the entire system.}
\label{fig:N1024_cgfield}
\end{figure}

We first focus on the case of a modest system size, $N = 1024$ ($L \sim 40$), and volume fraction, $\phi=0.7$. In \Fig{fig:N1024_cgfield}(a), we show the coarse-grained force field in a relaxed unjammed configuration (see \Sec{sec:model_method} for the definition). This shows that the field is correlated over the entire system, with a single region of weaker forces and a single domain where forces are larger. Both domains have a linear size comparable to $L$. 

We also show the coarse-grained field of the eigenmode associated with the lowest non-zero eigenfrequency $\omega_{\min}$ of the Hessian matrix for the same final configuration in \Fig{fig:N1024_cgfield}(b). These two fields are almost equivalent, confirming the deep connection between the softest vibrational mode in unjammed configurations and the final force field controlling the relaxation time $\tau$. This large correlation rationalises previous observations that $\tau \sim 1/\omega_{\min}^2$~\cite{Ikeda2020}. 

The visual impression is confirmed by a direct analysis of the projection of the eigenmode onto the force field. At this volume fraction, the projection yields $0.992(3)$. The equality $\tau = (2\omega_{\min}^2)^{-1}$ also holds numerically, confirming the harmonic nature of the final stage of the relaxation dynamics~\cite{Ikeda2020}.  

\begin{figure}
\includegraphics[width=8.5cm]{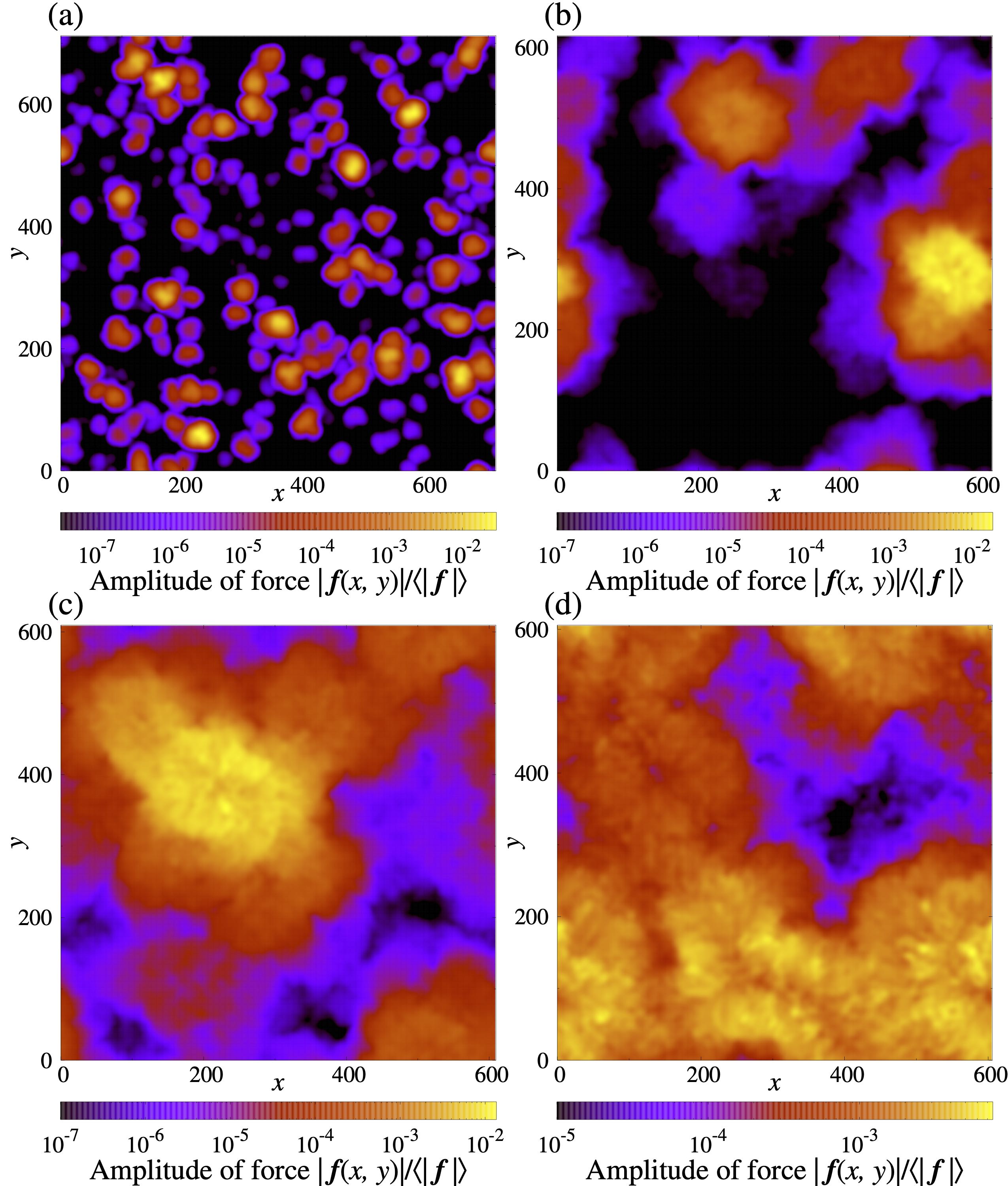}
\caption{Coarse-grained force field in $d=2$ for $N=262144$ and volume fractions (a) $0.6$, (b) $0.8$, (c) $0.82$, and (d) $0.83$. The size of the correlated large force `islands' (yellow) grows rapidly with increasing $\phi$.}
\label{fig:N262144_cgfield_density}
\end{figure}

The situation becomes very different with increasing the system size. \Fig{fig:N262144_cgfield_density} shows coarse-grained force fields
for $N=262144$ ($L \sim 600$) at various volume fractions. For $\phi=0.6$ and $\phi=0.8$, the force field does not form a single correlated region, but is instead composed of multiple `islands' where the force has a larger amplitude than in the rest of the system where forces are much smaller (the colour codes for forces in a logarithmic scale). The emergence of the multiple islands 
is unrelated to the connectivity percolation transition in the system, which occurs at $\phi \simeq 0.55$ \cite{Shen2012}. In our study, the system is always percolated and the multiple islands we identify with the force field all belong to the percolated cluster.
 
In such systems, we find that the softest eigenmode is mainly localized on a single one of these islands, while the other islands correspond to other eigenmodes with slightly larger frequencies. These observations were obtained by directly measuring the Hessian matrix in the final configuration for a system size $N=16384$, large enough to detect multiple islands and small enough that a diagonalisation of the Hessian remains numerically feasible. As a result, the projection of the softest eigenmode onto the residual force field decreases with increasing the system size.

The physical picture is that each of the large-force islands observed in the final configuration relaxes on its own timescale, but the relaxation time for the entire system is dominated by the slowest of these independently-relaxing islands. 

In addition, Fig.~\ref{fig:N262144_cgfield_density} shows that the typical size of these islands increases rapidly as the volume fraction increases towards jamming. As a result, a single correlated island seems to cover the whole system for $\phi = 0.83$, despite the fact that the system size is large $N = 262144$ (the linear size is larger than $L \sim 600$).

\begin{figure}
\includegraphics[width=\linewidth]{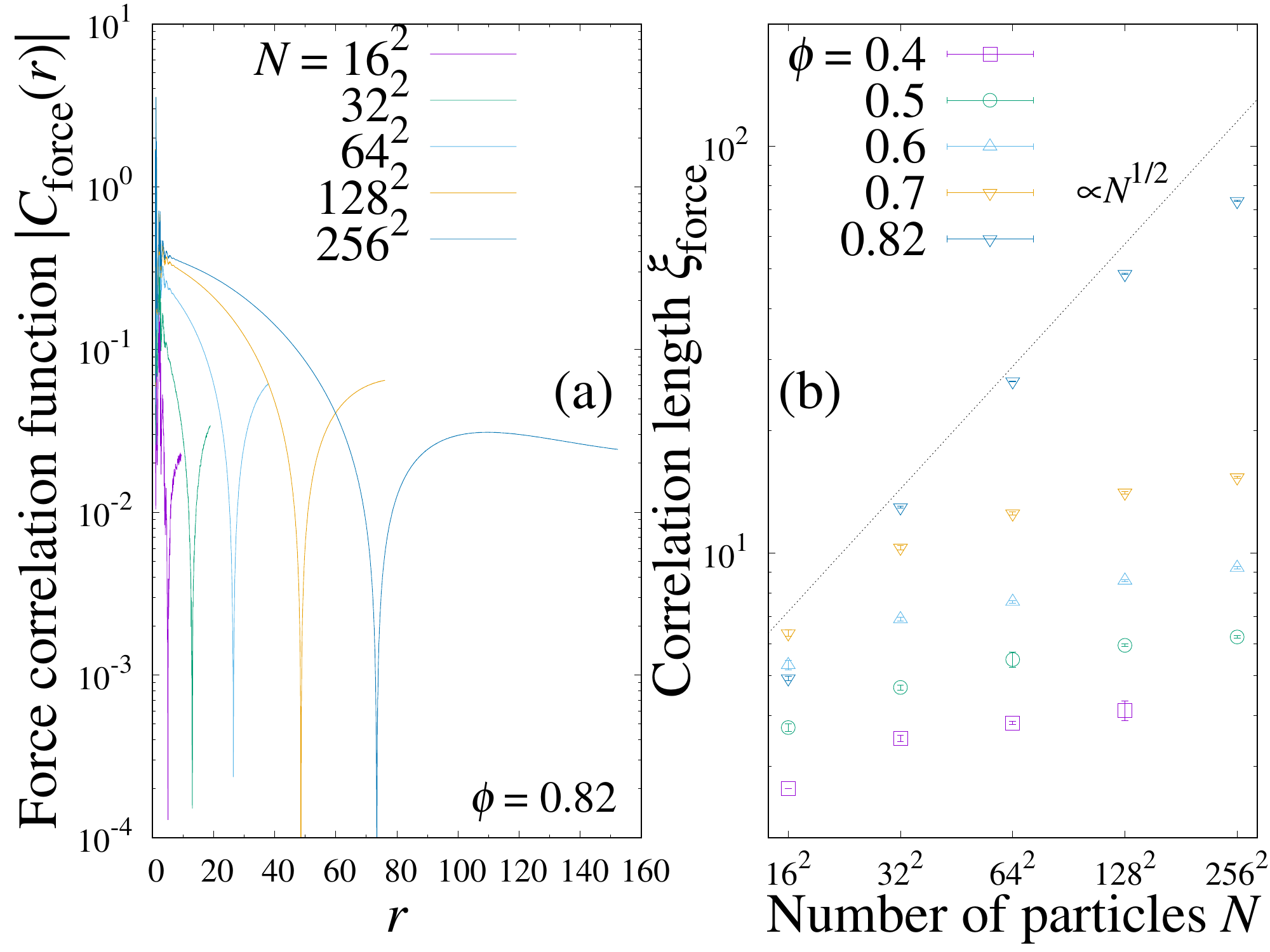}
\caption{(a) Absolute value of the force correlation function $|C_\mathrm{force}(r)|$ for $d=2$ and $\phi=0.82$ and various $N$.
The correlation length $\xi_\mathrm{force}$ is defined as the distance where $C_\mathrm{force}$ changes sign. (b) The evolution of $\xi_\mathrm{force}$ with $N$ for various $\phi$ changes from $\xi_\mathrm{force} \sim L \sim N^{1/d}$ to a slower logarithmic growth at large $N$.}
\label{fig:corr_length}
\end{figure}

To quantitatively analyze the emergence of the islands corresponding to domains where the force field is spatially correlated, we measure the spatial correlation function of the forces $C_{\rm force}(r)$, defined in Eq.~(\ref{eq:corr_func}). \Fig{fig:corr_length}(a) shows the absolute value $|C_{\rm force}(r)|$ for $\phi = 0.82$ and various system sizes. The absolute value is needed when using a logarithmic vertical axis, as the correlation changes sign at large $r$. We expect that $C_{\rm force}(r)$ first vanishes at a distance corresponding to the typical size of the islands seen in Fig.~\ref{fig:N262144_cgfield_density}. Therefore, we define the correlation length $\xi_\mathrm{force}$ as $C_{\rm force}(\xi_\mathrm{force}) = 0$ to measure the linear extension of the correlated force islands. 

The results are shown in \Fig{fig:corr_length}(b) which presents the evolution of $\xi_\mathrm{force}(\phi,N)$ for various values of $\phi$ and several system sizes $N$ in $d=2$. At low $\phi$, the correlation length $\xi_\mathrm{force}$ exhibits a mild increase with $N$ at large $N$ and its absolute value is modest. When $\phi$ increases (see for instance $\phi=0.82$), the growth of $\xi_\mathrm{force}$ is initially much stronger, compatible with $\xi_\mathrm{force} \sim L \sim N^{1/d}$. This suggests that in this regime, the force correlation length is actually bounded by the linear size of the system, compatible with the snapshots where the force field appears correlated over the entire system. For larger $N$, eventually, this very fast increase of 
$\xi_\mathrm{force}$ slows down and resembles the findings for low $\phi$, compatible with a slow, presumably logarithmic, growth. 

The data for the force correlation length suggest the existence of two regimes of system sizes, separated by a crossover length $\xi(\phi)$. There is a first regime at small $N$, i.e. $L < \xi(\phi)$, where the force correlation is limited by the system size, so that $\xi_\mathrm{force}(\phi,N) / L = O(1)$ and the system is composed of a unique correlated island. At larger $N$, i.e. $L > \xi(\phi)$, the force correlation enters a second regime where $\xi_\mathrm{force}(\phi,N) / L \ll 1$, and the system breaks into multiple independent islands. This behaviour echoes the evolution of the relaxation time which is nearly constant (or decreases slightly with $N$) in the first regime, and increases logarithmically in the second. Indeed, we observe that the typical system size where the crossover occurs, i.e. $L \simeq \xi(\phi)$, in both quantities is indeed similar.  

Importantly, the breaking of the system into independent sub-systems emerges when $N$ is large enough, but the crossover size $\xi(\phi)$ where this happens seems to depend very strongly on the volume fraction, and appears to become very large when the jamming transition is approached. The most natural interpretation is that the behaviour of all the quantities studied here is governed by a growing correlation lengthscale $\xi(\phi)$ which diverges as the jamming transition is approached. We discuss the physical content of $\xi(\phi)$ more extensively in Sec.~\ref{sec:summary}.
 
\subsection{Logarithmic growth explained by extreme-value statistics}

\label{sec:lg_evs}

We have established that the logarithmic growth of the relaxation time takes place in the regime where $L > \xi(\phi)$, when the system is large enough to exhibit multiple correlated islands where the relaxation dynamics can take place independently. We now use an extreme-value statistics argument to explain the logarithmic growth of $\tau$ with $N$.

\begin{figure}
\includegraphics[width=\linewidth]{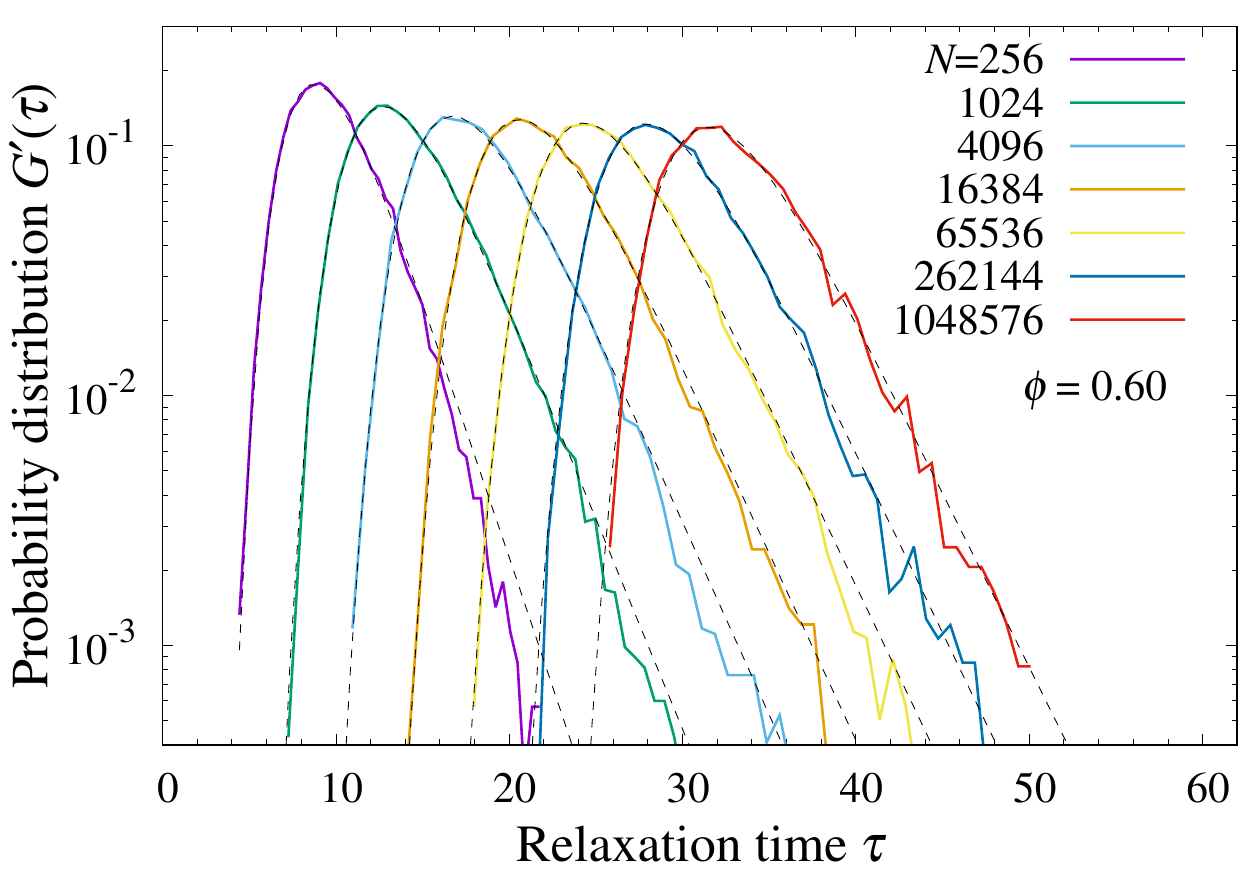}
\caption{Measured probability distribution functions of the relaxation time $\tau$ at $\phi = 0.6$ for several system sizes in $d=2$. Fits to \eq{eq:extreme-stat_CDF} are shown with dashed lines.}
\label{fig:extreme_arg}
\end{figure}

Let us suppose that the timescale $\tau_\mathrm{isl}$ controlling the relaxation locally in each island follows the probability distribution $P(\tau_\mathrm{isl})$. The global relaxation time of the system $\tau$ corresponds to the slowest timescale in a given configuration. We suppose that one configuration with $N$ particles can be decomposed into $M$ independent islands, which provide $M$ independent timescales $\{\tau_\mathrm{isl}^{(i)} \}_{i=1,\cdots,M}$. The probability that the maximum value in the set  $\{\tau_\mathrm{isl}^{(i)} \}$ is smaller than $t$ is $\mathrm{Prob}(t > \max \{\tau_\mathrm{isl}^{(i)} \}) = F(t)^M$, where $F(t) = \int_0^t P(\tau^\prime) d\tau^\prime$ is the cumulative distribution function of $P(\tau)$. The probability distribution for the largest time among the $M$ islands is thus $\partial (F(t)^M) / \partial t$. Assuming a simple form for the probability distribution $P(\tau) = \theta^{-1}\exp(-\tau/\theta)$, the average value $\langle \tau \rangle$ and the cumulative distribution $G(\tau)$ of the largest timescale, i.e. of the relaxation time $\tau$, are given by
\begin{align}
\label{eq:extreme-stat_ave}
\langle \tau \rangle 
&= M \int_0^\infty t F(t)^{M-1}P(t) dt
= \theta\sum_{k=1}^{M} \frac1k \sim \theta \log M,
\\
\label{eq:extreme-stat_CDF}
G(\tau) &= M \int _0^\tau F(t)^{M-1} P(t) dt = \bigpar{1 - \exp\bigpar{-\tau/\theta}}^M.
\end{align}
These expressions explain our numerical results quite well. To show this, we plot the probability distribution function of the relaxation time $G^\prime(\tau)$ obtained from numerical simulations in $d=2$ for $\phi = 0.6$ in \Fig{fig:extreme_arg}. Each of the measured distribution is fitted to the expression $G^\prime(\tau)=M\exp(-\tau/\theta)\bigpar{1 - \exp\bigpar{-\tau/\theta}}^{M-1}/\theta$, using $M$ and $\theta$ as fitting parameters. The fit is clearly excellent, using a nearly constant value $\theta \approx 2.9$, and a parameter $M$ growing rapidly with $N$ (a numerical fit gives $M \sim N^{0.77}$, close to the naive linear growth expected for the number of independent islands). This agreement supports the above argument using extreme-value statistics. 

The logarithmic growth of the relaxation time then follows from Eq.~(\ref{eq:extreme-stat_ave}), which shows that the largest relaxation time among $M$ islands scales as the logarithm of the number of islands $M$. A similarly slow growth would hold for any functional form of the distribution $P(\tau)$ provided its first moment is not divergent~\cite{Bardou2002,Fortin2015}. This shows that the emergence of multiple finite-sized islands in unjammed configurations is responsible for the logarithmic divergence of the relaxation time in the thermodynamic limit which eventually prevents the definition of the critical exponent $\beta$ from the relaxation time $\tau$. 

\section{Decoupling of shear viscosity and relaxation time}

\label{sec:shear}

Previous sections have shown that $\tau \sim (\Delta z)^{-\beta}$ in Eq.~(\ref{eq:critical_div}) cannot be correct as both $\tau$ and $\omega_{\min}$ are not defined in the unjammed phase in the thermodynamic limit. These two quantities are controlled by the slowest region of the entire system, whose timescale and lengthscale increases logarithmically with $N$.   

Regarding the shear viscosity, we can think of two possibilities: It either diverges like the relaxation time and it is not defined in the thermodynamic limit, or it remains finite and diverges as jamming is approached with a critical exponent $\beta$. The shear rheology is characterized by the steady state shear viscosity $\eta = \sigma / \dot \gamma$ and its pressure analogue $\eta_p = p / \dot \gamma$, where $\sigma$ is the shear stress, $\dot \gamma$ the shear rate, and $p$ the pressure. Previous studies suggested that the relaxation time $\tau$ measured after suddenly stopping the shear is proportional to the shear viscosity~\cite{Olsson2015}. This coupling was physically rationalised by the observation that the softest eigenmode of the Hessian controls both the relaxation dynamics~\cite{Ikeda2020} and the response to shear~\cite{Lerner2012}.

On the other hand, the above results for the relaxation dynamics from random configurations suggest that the situation may be different in large enough systems. Therefore, we need to revisit the relaxation dynamics from sheared configurations and its relation to the rheology. The corresponding simulation methods were described in Sec.~\ref{sec:model_method}. 

\begin{figure}
\includegraphics[width=\linewidth]{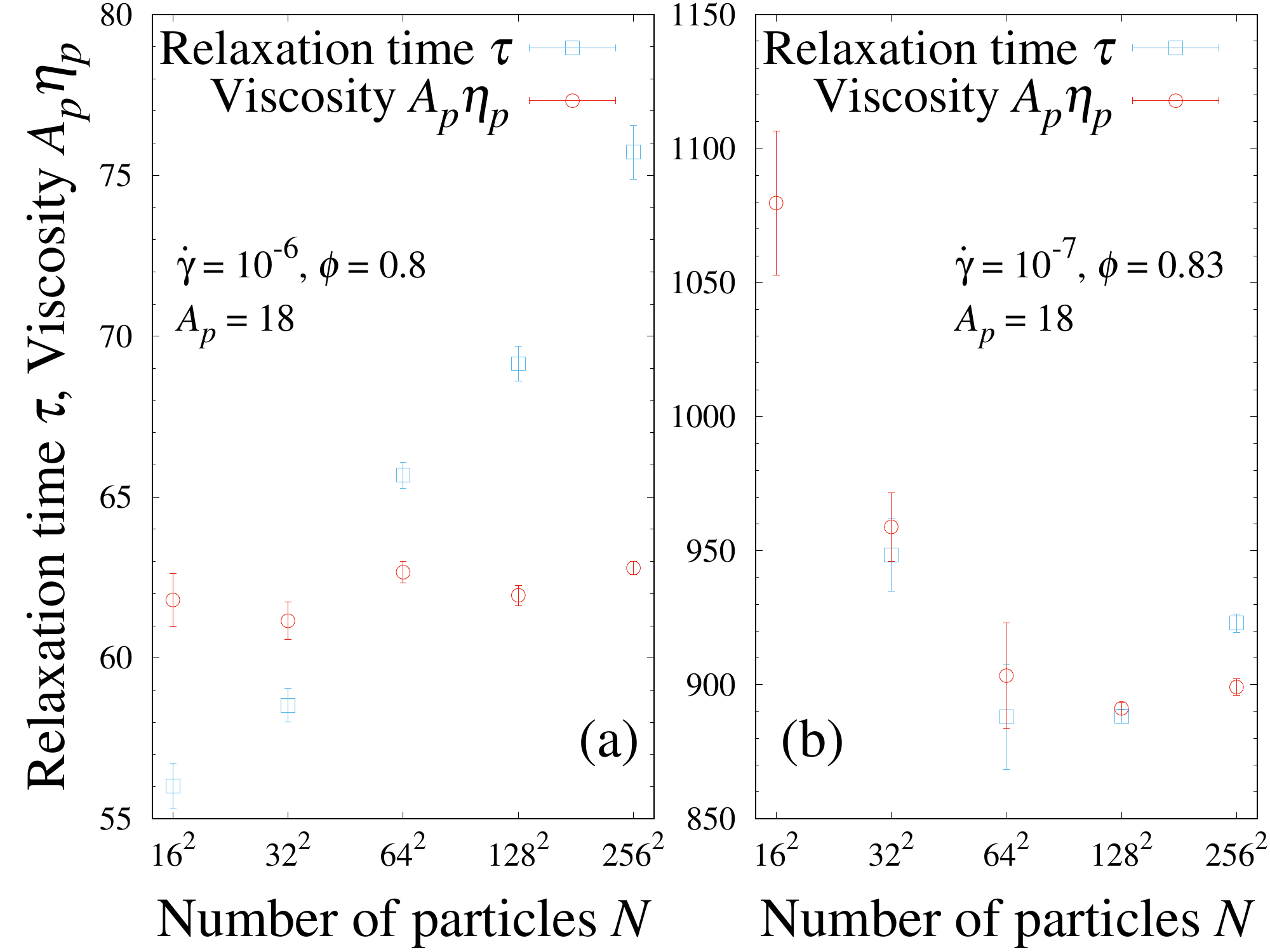}
\caption{Relaxation time and shear viscosity as a function of the system size for (a) $\phi = 0.8$ and (b) and $0.83$. 
We multiply $\eta_p$ with the constant $A_p = 18$~\cite{Olsson2015} for convenience. In (a), $\tau$ grows logarithmically with $N$, but the viscosity remains finite and is almost independent of $N$. In (b), both $\tau$ and $\eta_p$ first decrease with $N$, but only $\tau$ is expected to diverge as $N \to \infty$.}
\label{fig:tau_eta}
\end{figure}

Starting from sheared configurations in the steady state, we analyse the relaxation dynamics to measure the relaxation time $\tau$ as shown in  \Fig{fig:tau_eta}(a) for $\phi=0.8$ in $d=2$. Again, the relaxation time increases logarithmically with $N$, showing that it is divergent in the thermodynamic limit, just as for random configurations. \Fig{fig:tau_eta}(b) shows results for $\phi=0.83$, which mimic again the results for random configurations. Closer to jamming it becomes more difficult to observe the logarithmic behaviour at large $N$, as the crossover length $\xi(\phi)$ to enter the large $N$ regime is larger. 

We also find that the force field in the sheared case has the same properties as for the isotropic case. When the system size is small or the volume fraction is very close to jamming, the force field and the softest eigenmode are correlated over the whole system. These two fields are almost equivalent as they have a very large overlap. However, when the system size becomes large enough, multiple islands appear and the overlap between the two becomes small, even in the sheared case. As the particle configuration evolves with time during the steady shear, the relaxed configuration also changes depending on the starting configurations. The above findings about the islands hold very typically for the configurations in the steady states although we have not studied the detailed dynamics of the islands.

Next, we focus on the steady state shear viscosity. \Fig{fig:tau_eta}(a) compares the viscosity $\eta_p$ with the relaxation time $\tau$ for $\phi=0.80$. It is clear that the viscosity does not show the same logarithmic increase at large $N$ as $\tau$. We conclude that the shear viscosity is finite in the thermodynamic limit, and that it decouples from the relaxation time at large $N$. This implies that $\eta \sim \tau$ in Eq.~(\ref{eq:critical_div}) breaks down. 

We compare $\eta_p$ and $\tau$ much closer to jamming for $\phi=0.83$ in \Fig{fig:tau_eta}(b). In the regime of system sizes where the shear rheology can be analysed, the two quantities appear strongly coupled. For the largest $N$ value, we see a hint that $\tau$ enters the logarithmic regime whereas $\eta_p$ saturates to its large-$N$ limiting value, but this is difficult to see (despite the large system sizes studied).  

The physical interpretation of these results is that in the regime where $\xi_\mathrm{force} \sim L$, the coupling between viscosity and relaxation dynamics is strong and both quantities behave similarly. On the other hand, when $\xi_\mathrm{force} \ll L$, they become decoupled with $\tau(N) \sim \log N$ whereas $\eta_p$ saturates to a constant value. For a given volume fraction $\phi$, the decoupling thus occurs when $L > \xi(\phi)$, where $\xi(\phi)$ is the correlation length mentioned in Sec.~\ref{sec:island} above, and discussed further in Sec.~\ref{sec:summary}. 

The decoupling at large $N$ can be understood intuitively on the basis of the snapshots shown in Fig.~\ref{fig:N262144_cgfield_density}, which show the emergence of independent islands where the forces are large in unjammed configurations. Whereas the relaxation time is always dominated by the slowest of these multiple islands in each configuration, the viscosity presumably results from an ensemble average over all correlated domains. If the probability distribution of the local viscosity is well-behaved, its first moment has no reason to diverge in the thermodynamic limit. 

Note, however, that even though the viscosity saturates to a constant value at large $N$, it is subject to very strong finite-size effects which persist up to very large system sizes, $L \sim \xi(\phi)$. As noticed before, we expect these finite-size effects to become weaker if the  relaxation time and the viscosity are averaged at fixed contact number and not at fixed volume fraction, but the proper measurement of the viscosity requires $L \gg \xi(\phi)$.  

Several works have used scaling analysis to estimate the critical exponent of the shear viscosity~\cite{Olsson2011,Kawasaki2015,Olsson2020} and already pointed out that this is difficult due to a complicated scaling behavior near criticality and large corrections to scaling. Our results demonstrate that the large-$N$ limit of the viscosity is only accessible when the linear size of the system is much larger than a typical lengthscale $\xi(\phi)$ which seems to grow very fast as jamming is approached. For instance, a recent work estimated the $\beta = 2.68(8)$ in $d=2$ using a scaling analysis of the shear viscosity~\cite{Olsson2020}. However, the estimated exponent is very close to the apparent value for $\tau$ in Fig.~\ref{fig:tau_vs_dz}, and the range of system sizes used in Ref.~\cite{Olsson2020} belong to the regime $L < \xi(\phi)$. The large-$N$ limit for the shear viscosity appears even harder to achieve in larger dimensions.   

\section{Discussion}

\label{sec:summary}

In summary, we studied the relaxation dynamics of athermal frictionless soft spheres below jamming using extensive numerical simulations to directly test Eq.~(\ref{eq:critical_div}) and investigate the existence and numerical value of $\beta$ across different dimensions from $d=2$ to $d = \infty$. 

We discovered that the relaxation dynamics of unjammed packings close but below jamming is controlled by a large correlation lengthscale, $\xi(\phi)$, which diverges very fast as the jamming transition is approached. This finding, associated with the observation that two very different types of initial conditions yield similar results suggest that the dynamical slowing down approaching the jamming transition from below is largely universal. We expect that qualitatively similar results would be obtained using different types of local dynamics. Our results suggest that only collective algorithms relying on a detailed analysis of the microstructure of the force network could change the value of the dynamic critical exponents. 

For system sizes $L < \xi(\phi)$, the relaxation time, softest mode and shear viscosity are strongly coupled because the force field is correlated over the entire system and all these probes are physically equivalent. However, in this regime, all quantities are strongly affected by finite-size effects. 

In the other regime $L > \xi(\phi)$, the relaxation time and softest mode do not converge in the thermodynamic limit, because the system breaks into independently relaxing domains, and the relaxation time is dominated by the slowest region of the entire system. An extreme-value argument then explains its logarithmic divergence with system size, $\tau(N) \sim \log N$. In this regime, the exponent $\beta$ only applies to the shear viscosity which is then devoid of finite-size effects but this requires prohibitively large systems near jamming, making a precise determination of $\beta$ very difficult, even in $d=2$. 

The analysis of the mean-field Mari-Kurchan model suggests that the value $\beta \simeq 3.3$ should describe the large-$d$ limit. One may conjecture that this exponent describes the shear viscosity in any physical dimensions $d \geq 2$ in analogy with other jamming exponents. Due to the large finite-size effects mentioned above, we can neither rule out nor support this conjecture for the moment by direct numerical measurements. 

Because of the mean-field nature of the MK model, we expect that the mean-field dynamic exponent $\beta \simeq 3.3$ can be determined by analytic developments. Indeed, theoretical arguments developed for the sheared case give a quantitatively similar value $\beta \simeq 3.41$~\cite{DeGiuli2015}. In order to directly compare these two exponents, we need to study either the MK model under steady shear, or extend the theoretical argument to the isotropic case. The former is difficult because the random shifts couple particles that are spatially separated from each other by large distances, and a uniform shear flow is difficult to realize in the MK model. The latter seems more promising. We remark that Ref.~\cite{Hwang2020} recently proposed a variational argument for the dynamic exponent in the isotropic case which gives the same result as Ref.~\cite{Lerner2012a}, which is an earlier version of Ref.~\cite{DeGiuli2015}. We hope that our work will guide and inform further analytic developments. Indeed, since the initial submission of our manuscript, Ikeda~\cite{ikeda2020note} has obtained the same prediction $\beta=3.41$ for the isotropic case.  

\begin{figure}
\includegraphics[width=8.5cm]{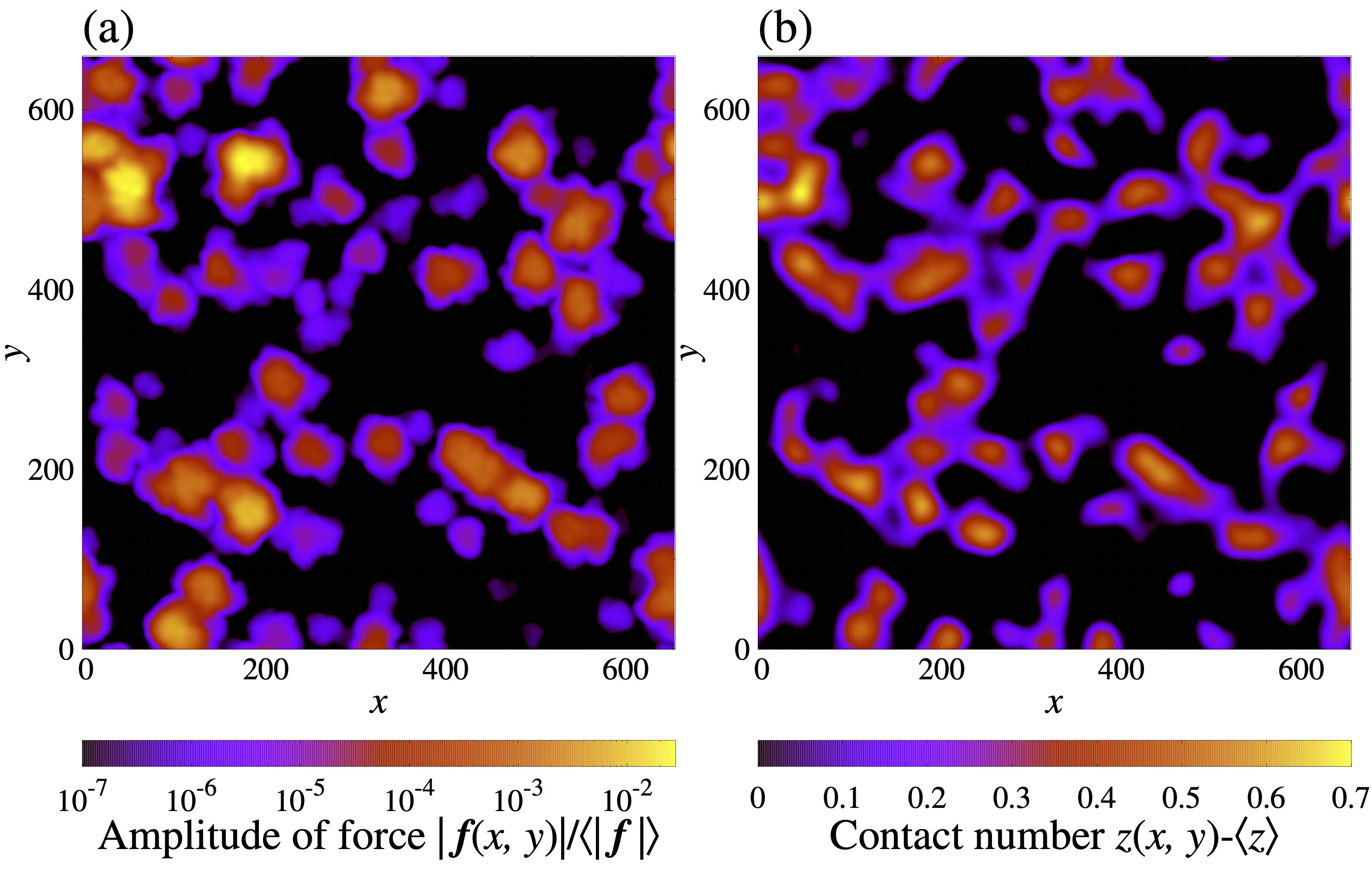}
\caption{Coarse-grained fields of (a) the amplitude of forces and (b) contact number fluctuations for $\phi = 0.7$ and $N=262144$. In (b), regions with negative $z(x,y) - \langle z \rangle$ are shown in black.}
\label{fig:N262144_fz_cgfield}
\end{figure}

What is the nature of the correlation length $\xi(\phi)$ controlling the two different regimes for the relaxation dynamics and shear rheology? The snapshots of the force field in Fig.~\ref{fig:N262144_cgfield_density} suggest that the microstructure of the system is very uniform when $L < \xi(\phi)$, and strongly heterogeneous when $L > \xi(\phi)$. The contact number $z$ also fluctuates spatially in unjammed packings, between regions that are highly connected and regions that are less connected. In \Fig{fig:N262144_fz_cgfield}, we compare the coarse-grained fields of the forces and of the contact number fluctuations, $\delta z({\vec r}) = z({\vec r}) - \langle z \rangle$, where $\langle z \rangle$ represents an average over the final configuration. This comparison confirms the intuition that the regions that are more connected are also the regions where the forces between particles are larger, and that the islands controlling the relaxation dynamics correspond to the well-connected regions. Recently, the spatial fluctuations of the contact number have been analysed numerically~\cite{Hexner2018,Hexner2019}. Combining our results with this recent analysis suggests to identify $\xi(\phi)$ with the correlation length of the contact number field. Numerically, a power law $\xi \sim (\Delta z)^{-\nu}$ was measured, with $\nu \simeq 0.7$ ($d=2$) and $\nu \simeq 0.85$ ($d=3$). This could also be consistent with a very recent study~\cite{Olsson2020} discussing the existence of a diverging length scale controlling the shear rheology of unjammed spheres, with a critical exponent close to $\nu = 1$. As reviewed in Ref.~\cite{Olsson2020}, the zoo of critical lengthscales near jamming is populated by many beasts, and it would be useful to provide a firmer theoretical basis for the correlation length $\xi(\phi)$ that controls the dynamics of unjammed packings. 

A final puzzle is the logarithmic increase of the force-force correlation length $\xi_{\rm force}$ in the regime $L > \xi(\phi)$, which appears incompatible, at first sight, with the multiple island picture given by the snapshots of the force field. Since the force-force correlation function \eq{eq:corr_func} is computed as an average over all particles, the correlation length $\xi_\mathrm{force}$ should indeed quantify the typical size of the islands. However, the amplitude of the forces in final configurations vary over orders of magnitude from one region to another, and we believe that here again the island with the largest forces in fact dominates the correlation function. The weak growth of the correlation length $\xi_\mathrm{force}$ in \Fig{fig:corr_length}(b) thus presumably results from a similar extreme-value mechanism as for the logarithmic growth of the relaxation time. 

\begin{acknowledgments}
We are grateful to M. Cates, H.~Ikeda, P~Olsson, M.~Wyart, and F. Zamponi for useful discussions. This work was supported by a grant from the Simons Foundation (Grant No. 454933, L. B.) and JSPS KAKENHI grants (No. 18H05225, 19H01812, 20H01868, and 20H00128, A. I.). 
\end{acknowledgments}

\bibliography{ref_jamm}

\end{document}